\documentclass[review]{article}
\usepackage{lineno, graphicx}

\newtheorem{thm}{Theorem}
\newtheorem{defin}{Definition}
\newtheorem{lemma}{Lemma}
\newtheorem{prop}{Proposition}

\def\squareforqed{\hbox{\rlap{$\sqcap$}$\sqcup$}}
\def\myqed{\ifmmode\squareforqed\else{\unskip\nobreak\hfil
\penalty50\hskip1em\null\nobreak\hfil\squareforqed
\parfillskip=0pt\finalhyphendemerits=0\endgraf}\fi}

\title{Tree spanners of small diameter\footnote{This result appears as a chapter in my PhD thesis at the
department of Computer Science, University of Toronto \cite{PhDthesis}}}

\author{
Ioannis Papoutsakis \\
Department of Informatics Engineering,\\
TEI of Crete, Heraklion, Crete, Greece.
}

\begin{document}
\maketitle

\begin{abstract}
A graph that contains a spanning tree of diameter at most $t$ clearly admits a tree $t$-spanner, since a tree $t$-spanner
 of a graph $G$ is a sub tree of $G$ such that the distance between pairs of vertices in the tree is at most $t$ times their
distance in $G$. In this paper, graphs that admit a tree $t$-spanner of diameter at most $t+1$ are studied. For $t$ equal to
1 or 2 the problem has been solved. For $t=3$ we present an algorithm that determines if a graph admits a tree 3-spanner
of diameter at most 4. For $t\geq4$ it is proved that it is an NP-complete problem to decide whether a graph admits a tree
$t$-spanner of diameter at most $t+1$.
\end{abstract}

\section{Introduction}

There are applications of spanners in a variety of areas, such as
distributed computing \cite{Awerbuch85,Peleg89},
communication networks \cite{PelUpf88,PelResh99}, motion planning and
robotics \cite{Arikati96,Chew89} and phylogenetic analysis
\cite{Bandelt86}. Furthermore, spanners are used in embedding finite
metric spaces in graphs approximately \cite{Rabinov98}.

On one hand, in \cite{Bondy89,CaiCor95a,CaiThesis} an efficient algorithm
to decide tree 2-spanner admissible graphs is presented, where a method
to construct all the tree 2-spanners of a graph is also given. On the
other hand, in \cite{CaiCor95a,CaiThesis} it is proven that for
each $t\geq 4$ the problem to decide graphs that admit a tree $t$-spanner
is an NP-complete problem. The complexity status of the tree 3-spanner
problem is unresolved. In this paper it is shown that the
problem to determine whether a graph admits a tree $t$-spanner of
diameter at most $t+1$ is tractable, when $t\leq 3$, while it is an
NP-complete problem, when $t\geq 4$.

Tree t-spanners ($t\geq 3$) have been studied for various families of
graphs. If a connected graph is a  cograph or a split graph or the
complement of a bipartite graph, then it admits a tree 3-spanner
\cite{CaiThesis}. Also, all convex bipartite graphs have a tree 3-spanner,
which can be constructed in linear time \cite{Venkatesan97}.
Efficient algorithms to recognize graphs that
admit a tree $3$-spanner have been developed for interval, permutation
and regular bipartite graphs \cite{Madanlal96}, planar graphs \cite{Fekete01},
directed path graphs \cite{Le99}, very strongly chordal graphs, 1-split graphs and
chordal graphs of diameter at most 2 \cite{Brandstadtchordal}. Moreover, every strongly chordal
graph admits a tree 4-spanner, which can be constructed in linear time
\cite{Brandst99}; note that, for each $t$, there is a connected chordal
graph that does not admit any tree $t$-spanner. In \cite{Brandstadtchordal} it is also presented
a linear time algorithm that finds a tree spanner in a small diameter chordal graph. In \cite{Manuel} the tree
$t$-spanner problem is studied for diametrically uniform graphs.

There are NP-completeness results for the tree $t$-spanner problem for families of graphs.
In \cite{Fekete01}, it is shown that it is NP-hard to determine the minimum $t$ for which a
planar graph admits a tree $t$-spanner. For any $t\geq4$, the tree $t$-spanner problem is NP-complete
on chordal graphs of diameter at most $t+1$, when $t$ is even, and of diameter at most $t+2$, when 
$t$ is odd \cite{Brandstadtchordal}; note that this refers to the diameter of the graph not to the diameter
of the spanner.

The tree 3-spanner problem is very interesting, since its complexity status is unresolved. In \cite{PhDthesis} it is
shown that only for t=3 the union of any two tree t-spanners of any given graph may contain big induced cycles but never
an odd induced cycle (other than a triangle); such unions are proved to be perfect graphs.
The tree 3-spanner problem can be formulated as an integer
programming optimization problem. Constraints for such a formulation appear in \cite{PhDthesis}, providing certificates
for non tree 3-spanner admissibility.

\section{Definitions and lemmas}
The definition of
tree a $t$-spanner follows, while, in general, terminology of \cite{West} is used.
\begin{defin}
A graph $T$ is a tree $t$-spanner of a graph $G$ if and only if $T$ is a
subgraph of $G$ that is a tree and, for every pair $u$ and $v$ of vertices of
$G$, if $u$ and $v$ are at distance $d$ from each other in $G$, then $u$
and $v$ are at distance at most $t\cdot d$ from each other in $T$.
\end{defin}

Note that in order to check that a spanning tree of a graph $G$
is a tree t-spanner of $G$, it suffices to examine pairs of adjacent
in $G$ vertices. A path of even length has a central vertex, while a path of odd length
has a central edge. We take into account this parity fact to define a
class of trivially tree $t$-spanner admissible graphs.
\begin{defin}
A $t$-center $K$ of a graph $G$ is a subgraph
of $G$ consisting exactly of either a vertex  when $t$ is even, or a pair of
adjacent in $K$ vertices when $t$ is odd, such that for all
$u$ in $G$, $d_G(K,u)\leq \lfloor \frac{t}{2} \rfloor$.
\end{defin}
Clearly, for any $t$-center $K$, we see that $|K|=|E(K)|= t \bmod 2$. 
Assume that a graph $G$ contains a $t$-center $K$. Any  Breadth-First Search
tree $T$ of $G$ starting from $K$ has the property that $d_T(K,u)=d_G(K,u)$,
for every vertex $u$ of $G$ \cite{West}. Therefore, since $K$ is a $t$-center
of $G$, $T$ is a tree $t$-spanner of $G$; observe that the distance in
$T$ between any pair of vertices $u$ and $v$ is at most equal to the
distance from $u$ to $K$ plus $|K|$ plus the distance from $K$ to $v$.
Graphs that admit a $t$-center are defined to be $t$-stars:
\begin{defin}
A graph $G$ is a $t$-star if and only if $G$ is the one
vertex graph or $G$ admits a $t$-center.
\end{defin}

Note that the one vertex graph is a $(2t+1)$-star for any $t$,
but it does not admit a $(2t+1)$-center. Also, observe that the only
0-star is the one vertex graph and the only 1-stars are the one vertex
graph and the one edge graph. If $k\leq t$, then a $k$-star is also a
$t$-star. Moreover, a $t$-star is a connected graph. Also, every connected
graph is a $t$-star, for some $t$. For example, a path of length $t$ is a
$t$-star and has a unique $t$-center. If a graph is a $t$-star, then
at least one spanning tree of the graph has diameter at most $t$:

\begin{lemma}
A graph $G$ is a $t$-star if and only if $G$ admits a spanning tree
of diameter at most $t$.
\label{ldiametros}
\end{lemma}

{\em Proof.}
On one hand, assume that $G$ is a $t$-star. If $G$ is the one
vertex graph, then the graph itself is such a spanning tree. Otherwise,
let $K$ be a $t$-center of $G$. Then, as we mentioned earlier, any
Breadth-First Search tree of $G$ starting from $K$ has diameter at most $t$.

On the other hand, assume that $G$ is not a $t$-star for some $t$.
We prove that the diameter of any spanning tree of $G$ is strictly
greater than $t$. Towards a contradiction, assume that $G$ admits a
spanning tree $T$ of diameter $d$ such that $d\leq t$. 
Consider a longest path $P$ of $T$, then $d=|P|$. Also, $P$ itself
is a $d$-star with a unique $d$-center $K$. Let $P_1$ and $P_2$
be the subpaths of $P$ from $K$ to $u$ and from $K$ to $v$, respectively,
where $u$ and $v$ are the endpoints of $P$. Then,
$|P_1|=|P_2|=\lfloor \frac{d}{2} \rfloor$.

Let $x$ be an arbitrary vertex of $G$. Since $T$ is a spanning tree
of $G$, there is a path $P'$ from $x$ to $K$ in $T$. Path $P'$ cannot
intersect (out of $K$) with both of $P_1$ and $P_2$, because otherwise
$T$ contains a cycle. So, we may assume that $P'$, $P_1$ and $K$ form
a path and $d_T(x,u) = |P'| + |P_1| + |K|$. But $d_T(x,u)\leq d$, since $d$
is the diameter of $T$. Therefore,
$d_T(K,x) = |P'| = d_T(x,u) - |P_1| - |K| \leq \lfloor\frac{d}{2}\rfloor$. 

But for every vertex $x\in G$, $d_G(K,x)\leq d_T(K,x)$.
Also, $\lfloor \frac{d}{2} \rfloor \leq \lfloor \frac{t}{2} \rfloor$,
since $d\leq t$. So $K$ is a $t$-center of $G$, which is a contradiction.\myqed

As a corollary, a tree is a $t$-star if and only if it has diameter at most
$t$, because a tree has only one spanning tree.
Now, for general graphs, if a graph is a $t$-star, then the diameter of the
graph is at most $t$ but there are graphs of diameter $d$ that are not
$d$-stars; for example, a cycle on eight vertices has diameter 4, but it
is not a 4-star.

When we examine a graph which is not a $t$-star, we actually face the 
tree $t$-spanner problem. Before presenting a frequently used lemma,
we give a definition to handle long paths.

\begin{defin}
A $t-midst$ $M(P,t)$ of a path $P$ from $u$ to $v$ is a subpath of
$P$ consisting exactly of either one vertex when $t$ is odd, or a pair
of adjacent in $M$ vertices when $t$ is even, such that
$d_P(M,u) > \lfloor \frac{t-1}{2} \rfloor$ and
$d_P(M,v)>\lfloor \frac{t-1}{2} \rfloor$.
\end{defin}

Obviously, a path $P$ has a $t$-midst if and only if $|P|\geq t+1$.
There may be many $t$-midsts in a path but only if $|P|=t+1$, does $P$ have
a unique $t$-midst. Clearly, for any $t$-midst $M$, we see that
$|M|=|E(M)|=(t+1) \bmod 2$.

\begin{figure}[htbp]
\begin{center}
\includegraphics{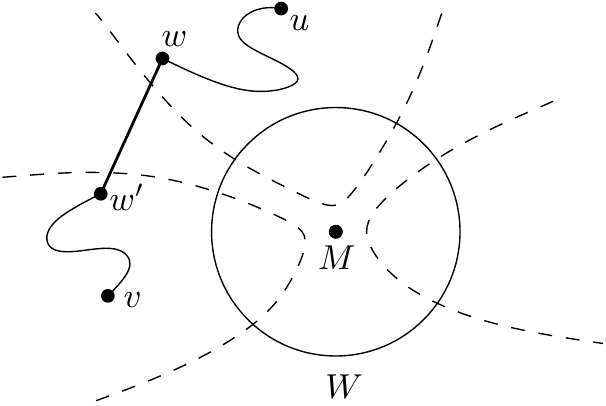}
\caption{Let $W$ (solid circle) be the vertices of $G$ within distance
$\lfloor \frac{t-1}{2} \rfloor$ from $t$-midst $M$ in $T$. How are the
components of $T\setminus M$ (dashed lines) related to the components
of $G\setminus W$? Here, $t$ is odd.}
\label{fnpalikari}
\end{center}
\end{figure}

\begin{lemma}
Let $G$ be a graph and $T$ a tree $t$-spanner of $G$. If $M$ is a
$t$-midst of a $u,\,v$-path $P$ of $T$, then every $u,\,v$-path $P'$ of
$G$ contains a vertex whose distance from $M$ in $T$ is at most
$\lfloor \frac{t-1}{2} \rfloor$.
\label{lofthalmos}
\end{lemma}
{\em Proof.}
{\em Proof.}
Since $G$ contains a path that admits a $t$-midst, $t$ is not zero.
Let $x=M$, when $t$ is odd, and $x=E(M)$ when $t$ is even.
Consider the components of $T\setminus x$. Note that when $t$ is even,
only two components are formed. Obviously, vertices $u$ and $v$ belong
to different such components. Therefore, for any $u,\,v$-path $P'$ of $G$
there is an edge $ww'$ in $P'$ such that $w$ is in a different component
than $w'$ (see figure~\ref{fnpalikari}).
Since all the tree paths connecting vertices of different such components
pass through $x$, it holds that $d_T(w,w')=d_T(w,M)+d_T(M,w')+|M|$.
But the tree distance between $w$ and $w'$ can be at most
$t$, therefore at least one of $w$ or $w'$ is within distance
$\lfloor \frac{t-1}{2} \rfloor$ from $M$ (consider different cases
when $t$ is odd or even; note that when $t$ is even the edge of $M$
participates in the tree path between vertices $w$ and $w'$). \myqed

\section{Overview}
Graphs that admit a tree $t$-spanner of small diameter are the
subject of this paper. First, for a $t$-star $G$, recall that any shortest
paths spanning tree to a $t$-center of $G$ is a tree $t$-spanner of $G$.
Second, according to theorem~\ref{tisesapostaseis}, if a graph admits a tree
$t$-spanner of diameter at most $t+1$, then at least one of its shortest
paths spanning trees is a tree $t$-spanner of the graph. Third, for each
$t\geq2$, there are graphs which admit a tree $t$-spanner of diameter
$t+2$ but none of their tree $t$-spanners is a shortest paths spanning tree (section~\ref{selxapostdendra}).
Therefore, should we expect that for each $t$ there is an efficient
algorithm to determine if a graph admits a tree $t$-spanner of diameter
at most $t+1$?

Theorem~\ref{tgiaolatat} settles this question. Consider a graph $G$ that
admits a tree $t$-spanner of diameter $t+1$ and let $K$ be the $(t+1)$-center of the tree. Then,
from theorem~\ref{tisesapostaseis}
we know that $G$ admits a tree 3-spanner $T$ which is a shortest paths
to $K$ spanning tree of $G$. When $t$ is odd, $K$ is just a vertex, so
there is a vertex $u$ in $G$ such that all the edges of $G$ incident to
$u$ are in $T$.  For $t=3$, finding the remaining edges of such a tree
3-spanner can be done efficiently (proposition~\ref{t3spandiam4}). Though, for each $t\geq 4$, the
problem of determining if a graph admits a tree $t$-spanner of diameter
at most $t+1$ is an NP-complete problem, where we use a reduction from
3-SAT problem. Note that the situation for this problem from
the complexity point of view is the same as the situation for the
standard tree $t$-spanner problem, except for the $t=3$ case for which the complexity
status of the standard problem is unresolved.

\section{Shortest paths spanning trees}
\label{selxapostdendra}
Let $K$ be a subtree of a graph $G$; in this paper $K$ is just a vertex or a
pair of adjacent vertices.
A shortest paths to $K$ spanning tree $T$ of a graph $G$ is a spanning tree
of $G$, such that for every vertex $x$ in $G$ the
unique path of $T$ between $x$ and $K$ is one of the shortest paths of
$G$ between $x$ and $K$. For every $d>0$, recall that a tree has a $d$-center if and only
if it is of diameter at most $d$ (lemma~\ref{ldiametros}).

\begin{thm}
If a graph $G$ admits a tree $t$-spanner which has a
$(t+1)$-center $K$, then $G$ admits a tree $t$-spanner $T$, such that
$d_G(K,x)=d_T(K,x)$, for every vertex $x$ in $G$.
\label{tisesapostaseis}
\end{thm}

{\em Proof.}
Let $K$ be a $(t+1)$-center of a tree $t$-spanner of a graph $G$.
Amongst all the tree $t$-spanners of $G$ with $(t+1)$-center $K$
(we know that there is at least one such tree $t$-spanner of $G$),
let $T$ be one such that the number of vertices of $G$ for which the
condition of the theorem does not hold is minimized and let $X$ be this
set of vertices, i.e. $X$ is the set of vertices $x$ of $G$ for which
$d_G(K,x)<d_T(K,x)$. Towards a contradiction assume that $X$ is nonempty.
Since $G$ is connected, there is a shortest path $P$ in $G$ from a vertex in
$X$ to $K$. Since $K$ does not have any vertex in common with $X$, path $P$
contains two consecutive in $P$ vertices $u$ and $v$, such that $u$ is in
$X$ and $v$ is not in $X$. As $P$ is a shortest path to $K$,
$d_G(K,u)=d_T(K,v)+1$, since $d_G(K,v)=d_T(K,v)$. Also, edge $uv$ is not in
$T$, because otherwise $u$ would not be in $X$. Let $e$ be the edge of $T$
incident to $u$ towards $K$. The tree path between $u$ and $v$ contains
edge $e$, because otherwise the tree path from $v$ to $K$ would contain
$u$. So, if we replace edge $e$ of $T$ with edge $uv$, the result
is another spanning tree $T'$ of $G$. Note that $K$ is a $(t+1)$-center
of both $T'$ and $T$, since $d_{T'}(K,x)\leq$
$d_T(K,x)\leq$ $\lfloor \frac{t+1}{2}\rfloor$ for every vertex $x$ of $T'$.

We prove that $T'$ is a tree $t$-spanner of $G$. It suffices to examine
the vertices in the component $Q$ of $T\setminus\{e\}$ that contains vertex
$u$, because the distances in $T'$ amongst the remaining vertices are the
same as in $T$. So, let $q$ be an arbitrary vertex of $Q$. Vertex $q$ is
within distance $\lfloor\frac{t+1}{2}\rfloor$ from $K$ in $T$, because
$K$ is a $(t+1)$-center of $T$. So,
$d_{T'}(q,K)\leq\lfloor\frac{t+1}{2}\rfloor-1$, since the $T'$ path from $q$
to $K$ contains vertex $u$ and $u$ is strictly closer to $K$ in $T'$ than 
it is in $T$. Now, for every vertex $p$ of $G$,
$d_{T'}(q,p)\leq$ $d_{T'}(q,K)+|K|+d_{T'}(K,p)\leq$
$\lfloor \frac{t+1}{2}\rfloor -1 + |K| + \lfloor \frac{t+1}{2}\rfloor =t$
(note that $|K|=1$, when $t+1$ is odd, and $|K|=0$, otherwise; also,
$d_{T'}(K,p)\leq \lfloor \frac{t+1}{2}\rfloor$, since $K$ is a $(t+1)$-center
of $T'$). Therefore, $T'$ is a tree $t$-spanner of $G$ with $(t+1)$-center
$K$. The fact that $d_{T'}(K,u)=d_G(K,u)$ for vertex $u$ is a contradiction
to the minimality of set $X$, since all the vertices of $T$ that are not
in $X$ satisfy the condition of the theorem for tree $T'$. Hence, $X$ is the
empty set, i.e.  $d_G(K,x)=d_T(K,x)$ for every vertex $x$ of $G$. \myqed

When we consider tree $t$-spanners with a $(t+2)$-center, this theorem does
not hold in general, except for the trivial case where $t\leq 1$. For each
$t\geq 2$, we now present an example of a  2-connected graph that admits a
tree $t$-spanner with a $(t+2)$-center and there is no tree $t$-spanner
of the graph which satisfies the condition of the theorem.

\begin{figure}[htbp]
\begin{center}
\includegraphics{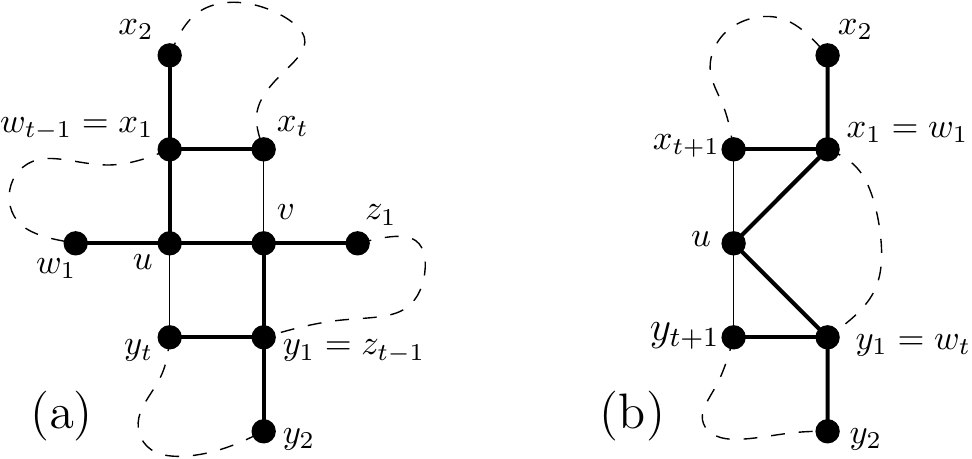}
\caption{(a) For each odd $t\geq3$, a graph that admits a tree $t$-spanner
with a $(t+2)$-center, where the dashed lines represent paths of length
$t-2$; when $t=3$, all the vertices of the graph are shown in the figure
and the bold edges form a tree 3-spanner of the graph with 5-center the
pair $u$ and $v$. (b) For each even $t\geq2$, a graph that admits a tree
$t$-spanner with a $(t+2)$-center, where the dashed lines represent paths
of length $t-1$; when $t=2$, all the vertices of the graph are shown in
the figure and the bold edges form a tree 2-spanner of the graph with
4-center $u$.}
\label{foxielaxapost}
\end{center}
\end{figure}

When $t$ is an odd number greater or equal to 3, consider a graph $G$
consisting of (1) two cycles $C_1=v,\,u,\,x_1,\ldots,\,x_t$ and
$C_2=u,\,v,\,y_1,\ldots,\,y_t$ each of
length $t+2$, such that $C_1\cap C_2=\{u,\,v\}$, (2) two edges $x_1x_{t}$,
$y_1y_{t}$ and (3) two paths $u,\,w_1,\ldots,\,w_{t-1}$=$x_1$ and
$v,\,z_1,\ldots,\,z_{t-1}$=$y_1$ each of length $t-1$ and
vertex disjoint (but their endpoints) from each other and from the two cycles
(see figure~\ref{foxielaxapost} (a)). If we remove
edges $vx_{t}$, $uy_{t}$, $x_{\lceil t/2\rceil}x_{\lceil t/2\rceil+1}$,
$y_{\lceil t/2\rceil}y_{\lceil t/2\rceil+1}$,
$w_{\lfloor t/2\rfloor}w_{\lceil t/2\rceil}$,
$z_{\lfloor t/2\rfloor}z_{\lceil t/2\rceil}$ from $G$, then the result is a tree
$t$-spanner $T$ of $G$, where the pair $u$ and $v$ induces a $(t+2)$-center $K$
of $T$. Now, edges $x_1x_{t}$ and $ux_1$ have to be in every tree $t$-spanner
of $G$, because, otherwise, if edge $x_1x_{t}$, for example, is not in a tree
$t$-spanner of $G$, then we are left with the edges of cycle $C_1$ which
has length $t+2$. So, vertex $x_t$ is at distance 2 from $K$ in all the
tree $t$-spanners
of $G$ having $K$ as a $(t+2)$-center, although $x_t$ is adjacent to $K$
in $G$, i.e. $x_t$ violates the condition of the theorem. Note that $G$ does
not contain any tree $t$-spanner with a $(t+1)$-center and $G$ is
2-connected. Also, for any arbitrary tree $t$-spanner of $G$ with a
$(t+2)$-center different than $K$ there is a vertex that violates
the condition of the theorem (for example, when $t\geq 5$, the pair $v$ and
$x_t$ can be a $(t+2)$-center of a tree $t$-spanner of $G$ and, then,
vertex $u$ violates the condition of the theorem).

When $t$ is an even number greater or equal to 2, consider a graph
$G$ consisting of (1) two cycles $C_1=u,\,x_1,\ldots,\,x_{t+1}$ and
$C_2=u,\,y_1,\ldots,\,y_{t+1}$ each of length $t+2$, such that
$C_1\cap C_2=\{u\}$, (2) two edges $x_1x_{t+1}$, $y_1y_{t+1}$
and (3) a path $x_1$=$w_1,\,\ldots,\,w_t$=$y_1$ of length $t-1$ (see
figure~\ref{foxielaxapost} (b)). If we remove edges
$ux_{t+1}$, $uy_{t+1}$, $x_{t/2+1}x_{t/2+2}$, $y_{t/2+1}y_{t/2+2}$
and $w_{t/2}w_{t/2+1}$ from $G$, then the result is a tree $t$-spanner $T$ of
$G$, where vertex $u$ is a $(t+2)$-center of $T$. Since edges $x_1x_{t+1}$
and $ux_1$ are $t$-forced, for every tree $t$-spanner of $G$, vertex $x_{t+1}$
is at distance 2 from $u$, although $x_{t+1}$ is adjacent to $u$ in $G$,
i.e. vertex $x_{t+1}$ violates the condition of the theorem.
Here, as well as when $t$ is odd, $G$ is a 2-connected graph and $G$
does not contain any tree $t$-spanner with a $(t+1)$-center.

\section{Tree $3$-spanners of diameter $4$}
When $t=2$, all the tree 2-spanners of a graph can be generated
using the skeleton tree of the graph~\cite{CaiCor95a,CaiThesis}; because
of that, one can efficiently decide if a graph admits a tree 2-spanner of
diameter at most $d$, for any $d$ (in case that the graph admits a skeleton
tree). For $t\geq3$, there is no tool known analogous to the skeleton tree.
According to theorem~\ref{tisesapostaseis},
if a graph admits a tree $t$-spanner with a $(t+1)$-center, then the graph
admits a tree $t$-spanner which is a shortest paths to this center spanning
tree of the graph. Because of this property, when $t=3$, there is a short
characterization of graphs that admit a tree 3-spanner of diameter at
most 4.

\begin{prop}
A graph $G$ admits a tree 3-spanner of diameter at most 4 if and
only if $G$ contains a vertex $u$, such that each component of
$G\setminus N[u]$ is included in the neighborhood of a neighbor
of $u$.
\label{t3spandiam4}
\end{prop}
{\em Proof.}
Assume that a graph $G$ contains a vertex $u$,
such that, for every component $Q$ of $G\setminus N[u]$, there is a vertex
$v_Q$ in $N(u)$ for which $V(Q)\subseteq N(v_Q)$. Consider the following
set of edges $E(T)=\{ux:\:x\in N(u)\}\cup\bigcup_{Q\in C} \{v_Qx:\:x\in Q\}$,
where $C$ is the set of components of $G\setminus N[u]$. Since every edge in
$E(T)$ belongs to $E(G)$, $E(T)$ induces a subgraph $T$ of $G$.
Every vertex of $G$ other than $u$ contributes exactly one edge (in
the direction to $u$) to $E(T)$ and
$E(T)$ does not contain any other edges; so, $E(T)$ contains $n-1$ edges.
Also, every vertex of $G$ other than $u$ is adjacent in $T$ either to $u$ or
to a neighbor of $u$ in $T$ and, therefore, $T$ is connected and has
diameter at
most 4. Thus, it suffices to prove that $T$ is a 3-spanner of $G$. First, if
a vertex is in $N_T(u)$, then it is within distance 3 in $T$ from 
every vertex of $G$, since $u$ is within distance 2 in $T$ from every vertex
of $G$. Second, for the remaining vertices, a vertex $x$ in a component $Q$
of $G\setminus N[u]$ is adjacent in $G$ either to a vertex in $N_T(u)$,
which we examined in the first case, or to a vertex $y$ which is also
in $Q$ and, therefore, $x$ and $y$ are at distance 2 from each
other in $T$ (edges $v_Qx$ and $v_Qy$ are both in $T$).

Assume now that a nonempty graph $G$ admits a tree 3-spanner $T'$ of diameter at most
4. Then, $T'$ has a $4$-center $u$ (even if $T'$ is the one vertex graph).
Because of theorem~\ref{tisesapostaseis}, $G$ admits a tree 3-spanner $T$,
such that $d_G(u,x)=d_T(u,x)$, for every vertex $x$ of $G$. Therefore,
$N_G(u)=N_T(u)$.
First, let $Q$ be any component of $G\setminus N_G[u]$. Since $u$ is a 4-center
of $T$, every vertex in $Q$ is adjacent in $T$ to some vertex in $N_G(u)$.
Assume that two distinct vertices $x$ and $y$ of $Q$ are adjacent in $T$
to two distinct vertices $v_x$ and $v_y$ of $N_G(u)$, respectively. Then,
$u$ is a 3-midst of path $x$, $v_x$, $u$, $v_y$, $y$ of $T$. Since $x$ and
$y$ are in the same component $Q$, there is a path in $Q$ between $x$
and $y$. But, none of the vertices of $Q$ is at distance 1 from $u$ in $T$,
which is a contradiction, because of lemma~\ref{lofthalmos}. Hence, all the
vertices in $Q$ are adjacent in $T$ to the same vertex, say $v_Q$, in $N_G(u)$.
So, for every component $Q$ of $G\setminus N_G[u]$, there is a vertex $v_Q$
in $N_G(u)$ for which $Q\subseteq N_T(v_Q)\subseteq N_G(v_Q)$.
Second, if $G\setminus N_G[u]$ is the empty graph, then there is no
component to examine and, therefore, the statement holds vacuously.\myqed

This proposition gives rise to an efficient algorithm to determine if
a graph admits such a tree 3-spanner. Assume we are given a graph $G$.
For every vertex $u$ of $G$, we can find efficiently the components of
$G\setminus N_G[u]$, using a breadth first search algorithm, for example.
It remains to check if for each such component $Q$ there is a neighbor
$v$ of $u$ for which $V(Q)\subseteq N(v)$, which can be easily performed.
The following program illustrates such an algorithm.

\begin{verbatim}
input(G);
flag=1;                 /*In case G is the empty graph
for u in V(G) {
   Let C be the set of components of G-N[u];
   flag=1;             /*In case G=N[u] and therefore C is empty
   for X in C {
      flag=0;
      for v in N(u)
         if (X is a subset of N(v)) {flag=1; break;}
      if (flag==0) break;
   }
   if (flag) break; else continue;
}
if (flag)
   output(G admits a tree 3-spanner of diameter at most 4);
else
   output(G does not admit a tree 3-spanner of diameter at most 4);
\end{verbatim}

Of course, if a graph admits a tree 3-spanner of diameter at most 4, i.e.
the tree 3-spanner is a 4-star, then the graph itself is a 4-star, as well.
In contrast,
it can be the case that a 4-star graph is tree 3-spanner admissible but
none of its tree 3-spanners has diameter at most 4. Figure~\ref{foxielaxapost}
(a) for $t=3$ depicts such a graph; vertex $u$ is a 4-center of the graph but
all the tree 3-spanners of this graph have diameter greater or equal to 5.

\section{Tree $t$-spanners of diameter $t+1$ for $t\geq 4$}
As shown in~\cite{CaiCor95a,CaiThesis} using a reduction from 3-SAT,
for each $t\geq 4$,
the problem of determining if a graph admits a tree $t$-spanner is an
NP-complete problem. In this reduction, for each instance of 3-SAT a graph is
generated and it is shown that an instance of 3-SAT is satisfiable if and only
if the corresponding graph is tree $t$-spanner admissible. It turns out
that if an instance of 3-SAT is satisfiable, then its corresponding graph
admits a tree $t$-spanner of diameter at most $2(t+\lceil\frac{t}{2}\rceil-1)$.
It is possible to alter slightly this reduction to prove that the problem of
determining if a graph admits a tree $t$-spanner of diameter at most
$2(t+\lceil\frac{t}{2}\rceil-3)$ is an NP-complete problem. For example,
the problem of determining if a graph admits a tree 4-spanner of diameter
at most 6 is an NP-complete problem. Unfortunately, it seems as though
the reduction used in~\cite{CaiCor95a,CaiThesis} cannot be employed for
lower diameters and we need to employ a much different reduction.

For each $t\geq 4$, we consider the following problem.
Given a graph $G$, does $G$ admit a tree $t$-spanner of diameter at most
$t+1$? Each of these problems belongs to NP, since, given graphs $G$ and $T$,
we can verify efficiently if $T$ is a tree $t$-spanner of $G$ of diameter at
most $t+1$. To see this, it suffices to verify that $T$ is a spanning tree of
$G$ of diameter at most $t+1$ and no pair of vertices being at distance $t+1$
apart in $T$ are adjacent in $G$.

We prove that each of these
problems is an NP-complete problem, using a reduction from 3-SAT.
An instance of 3-SAT consists of a set of clauses, where each clause
is the disjunction of exactly three distinct literals (see for example
~\cite{GarJon79}). A literal is a variable
or the negation of a variable. An instance of 3-SAT is satisfiable if and only
if there is a truth assignment to the variables that participate in the
clauses of the instance, such that all clauses of the instance become true.
The 3-SAT problem is to determine if such an instance is satisfiable.

\subsection{Stretch factor equals 4}
We present an algorithm $f$ which receives as input an instance $I$ of
3-SAT and outputs a graph $f(I)$. The basic steps of this algorithm
are the following.

(I) Given $I$, let $X(I)$ be the set of variables involved in $I$.
Each variable in $X(I)$ becomes a vertex of the
output graph $f(I)$. Also, consider a set of 6 vertices $H=\{u,$ $v$,
$h_u$, $h_u'$, $h_v$, $h_v'\}$, such that $H$ does not have any element
in common with $X(I)$.

(II) Let $C(I)$ be the set of clauses of $I$ which do not contain both
a variable and its negation, i.e. each clause in $C(I)$ contains exactly 3
distinct variables. For every clause $c$ in $C(I)$, we generate a
set of vertices $V^c$, such that  $V^c$ does not have any element in common
with $X(I)$ or $H$. Also, for every pair of clauses $c$ and $p$ in $C(I)$,
sets $V^c$ and $V^p$ do not have any vertex in common. To make this clear
we use $c$ as a superscript on each vertex of $V^c$. Let $c$ be a clause
in $C(I)$, which involves variables $x_1$, $x_2$ and $x_3$, say; then,
$V^c=\{x_1^c,\,x_2^c,\,x_3^c\}\cup\{q_1^c,\ldots,\,q_8^c\}$. Let
$Q(I)=\bigcup_{c\in C(I)} V^c$. Upon input $I$ consider the vertex set
$V(I)= H \cup X(I) \cup Q(I)$.

(III) We continue with the edge set of the output. Consider the following
matrix:
\[M=\left[ \begin{array}{cccccccc}
       1 & 1 & 1 & 1 & 0 & 0 & 0 & 0 \\
       0 & 0 & 0 & 0 & 1 & 1 & 1 & 1 \\
       1 & 1 & 0 & 0 & 1 & 1 & 0 & 0 \\
       0 & 0 & 1 & 1 & 0 & 0 & 1 & 1 \\
       1 & 0 & 1 & 0 & 1 & 0 & 1 & 0 \\
       0 & 1 & 0 & 1 & 0 & 1 & 0 & 1 \end{array} \right] \]
Matrix $M$ has two main properties. On one hand, it consists of three distinct
pairs of complementary to each other rows (for example the first row is the
complement of the second). On the other hand, if a sub-matrix of $M$
consisting of whole rows of $M$ contains at least one 1 in each column,
then the sub-matrix must contain at least one pair of complementary to each other
rows. Let $c$ be a clause of $C(I)$ that contains variables $x_1$, $x_2$ and
$x_3$, say, where $x_1$ appears first in $c$, $x_2$ appears second and $x_3$
appears third. Consider the ordering (vector) $g^c$=$[x_1$, $x_1^c$,
$x_2$, $x_2^c$, $x_3$, $x_3^c]$ of vertices of $f(I)$. We
generate a set of edges $E^c$ between vertices in $V(I)$ as follows. First,
for $1\leq i \leq 6$ and $1\leq j \leq 8$, if $M_{i,j}=1$, then edge $g^c_iq_j^c$
belongs to $E^c$, i.e. matrix $M$ becomes the adjacency matrix between
vertices in $g^c$ and vertices in $\{q_1^c,\ldots,q_8^c\}$. Second, for
$1\leq i \leq 3$, if variable $x_i$ appears positive in $c$, then edge
$x_i^cu$ belongs to $E^c$; else if the negation of $x_i$ appears in $c$, then
edge $x_i^cv$ belongs to $E^c$ (i.e. we use vertex $x_i^c$ to
indicate the standing of variable $x_i$ in $c$). Finally,
for $1\leq i \leq 7$, edge $q_iq_{i+1}$ belongs to $E^c$ and $E^c$ does not
contain any other edges. We consider the union of these edge sets over all
clauses $c$ of $C(I)$; so, let $E(Q)=\bigcup_{c\in C(I)}E^c$.

(IV) For vertices in $X(I)$ we consider the following set of edges, $E(X)$.
For every $x$ in $X(I)$, edges $xu$ and $xv$ belong to $E(X)$, and
$E(X)$ does not contain any other edges. Also, let
$E(H)=\{h_uh_u'$, $h_u'u$, $uv$, $vh_v'$, $h_v'h_v\}$, i.e. vertices in
$H$ form an $h_u,\,h_v$-path of length 5.
Finally, let $E(I)= E(Q)\cup E(X)\cup E(H)$.

(V) Algorithm $f$ outputs $f(I)=(V(I),E(I))$.

\begin{figure}[htbp]
\begin{center}
\includegraphics[width=8cm]{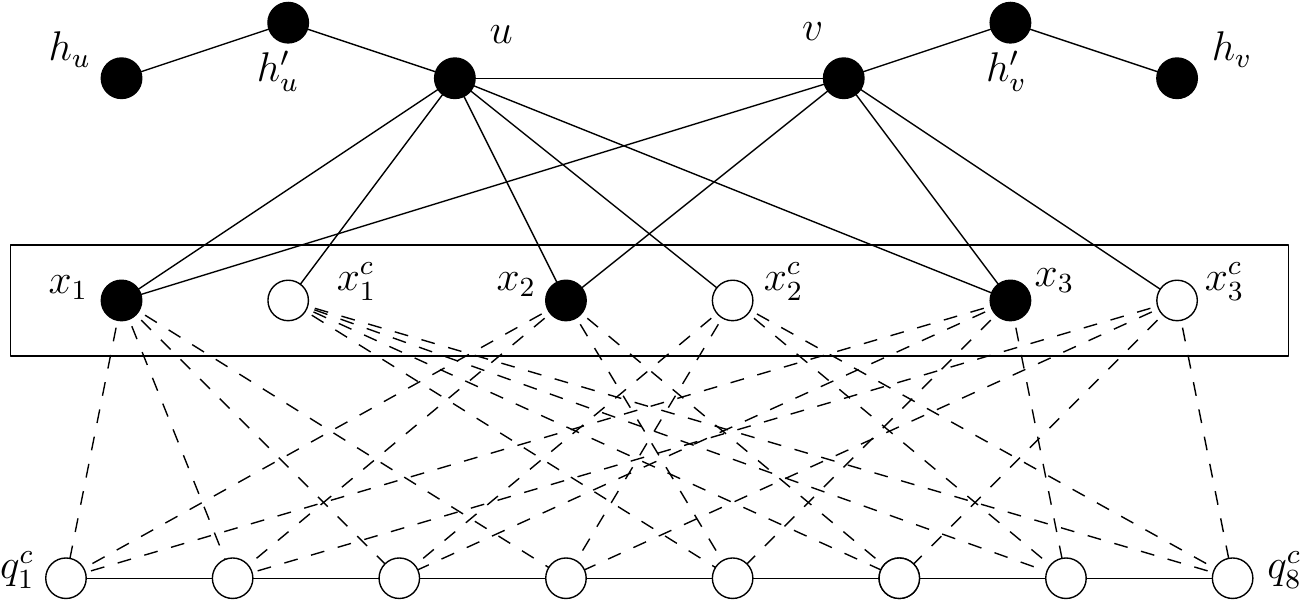}
\caption{The subgraph of $f(I)$ induced by vertices in $H\cup V^c\, \cup$
$\{x_1,\,x_2,\,x_3\}$, where $c$ is clause $(x_1\vee x_2 \vee \neg x_3)$
of $I$. Notice that $x_1^c$ and $x_2^c$ are adjacent to $u$ while $x_3^c$
is adjacent to $v$, since variables $x_1$ and $x_2$ appear positive in
$c$ while the negation of $x_3$ appears in $c$. The vertices in the
rectangle are the vertices in vector $g^c$ ordered from left to right,
where $g_1^c$ is the leftmost vertex in the rectangle, i.e. $g_1^c$ is
vertex $x_1$. The dashed lines represent the edges determined by matrix $M$.
Also, vertices in $V^c$ (white vertices) are incident in $f(I)$ only to edges
shown in the figure.}
\label{f4spandiam5}
\end{center}
\end{figure}

For every instance $I$ of 3-SAT, output $f(I)$ of algorithm $f$ is a
graph, since $E(I)$ contains edges with endpoints in $V(I)$. Note that
$f$ runs in polynomial time.
Figure~\ref{f4spandiam5} shows part of such an output. Note that
clauses that contain both a variable and its negation are disregarded
in the construction of $f(I)$.

\begin{prop}
For every instance $I$ of 3-SAT, $I$ is satisfiable if and only if graph
$f(I)$ admits a tree 4-spanner of diameter at most 5.
\label{pNP4diam5}
\end{prop}

{\em Proof.}
Let $A$ be a truth assignment that satisfies an instance
$I$ of 3-SAT. Then, for every clause $c$ of $I$, let $x(c)$ be one of
the variables that make clause $c$ true with respect to $A$.  

Consider the following set of edges, $T^X$. For every variable $x$ of $X(I)$,
if $A(x)$ is true, then edge $xu$ belongs to $T^X$, otherwise ($A(x)$ is false)
edge $xv$ belongs to $T^X$. Also, let $T^H=\{h_uh_u'$, $h_u'u$, $uv$,
$vh_v'$, $h_v'h_v\}$=$E(H)$.

Let $C(I)$ be the set of clauses of $I$ that contain 3 distinct variables.
For each clause $c$ of $C(I)$ which contains variables $x_1$, $x_2$ and $x_3$,
say, we consider the following set of edges $T^c$. For $1\leq j \leq 3$,
the edge of graph $f(I)$ from $x_j^c$ to $u$ or $v$ belongs to $T^c$
(note that each of $x_j^c$ is adjacent to exactly one of $u$ or $v$
in $f(I)$). Also, $x(c)=x_i$ for some $i$ from 1 to 3. Then, all
the edges of graph $f(I)$ between $\{x_i$, $x_i^c\}$ and
$\{q_1^c,\ldots,q_8^c\}$ belong to $T^c$. Finally, let
$T^Q=\bigcup_{c\in C(I)} T^c$.

\begin{figure}[htbp]
\begin{center}
\includegraphics[width=\textwidth]{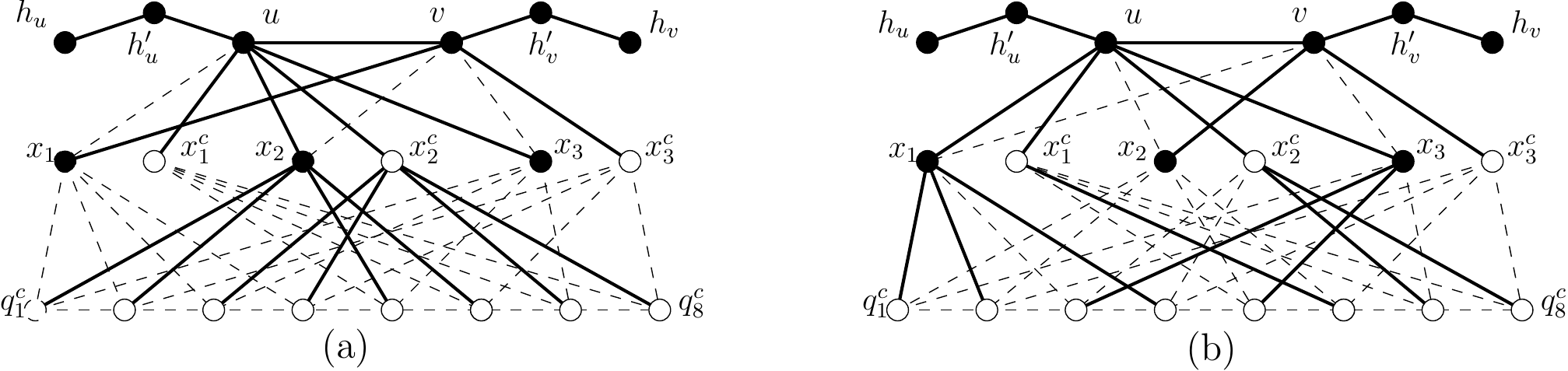}
\caption{Two copies of the subgraph of $f(I)$ induced by vertices in
$H\cup V^c\, \cup$ $\{x_1,\,x_2,\,x_3\}$, where $c$ is clause
$(x_1\vee x_2 \vee \neg x_3)$ of $I$. On the left, bold edges
represent edges of $T$, where $T$ is constructed upon a truth assignment
$A$ that satisfies $I$. Here, $A(x_1)$ is false, $A(x_2)$ is true and
$A(x_3)$ is true. There is only one choice for $x(c)$, namely $x(c)=x_2$.
Observe that both of $x_2$ and $x_2^c$ are adjacent in $T$ to $u$ and each
vertex in $\{q_1^c,\ldots,q_8^c\}$ is adjacent in $T$ to exactly one of $x_2$ or
$x_2^c$. On the right, bold edges are edges of a tree 4-spanner $T$ of $f(I)$
which is a shortest paths to $K$ spanning tree of $f(I)$. Given $T$, we
generate a truth assignment $A$ for which $A(x_1)$ is true, $A(x_2)$ is
false and $A(x_3)$ is true. Since both of $x_1$ and $x_1^c$ are adjacent to
$u$ in $T$, variable $x_1$ makes $c$ true. Observe that the first and
second rows of $M$ were needed to cover all the vertices in
$\{q_1^c,\ldots,q_8^c\}$.}
\label{fNP4diam5}
\end{center}
\end{figure}

We prove that the set of edges $T^X \cup T^H \cup T^Q$ induces a tree
4-spanner $T$ of $f(I)$ of diameter at most 5. Obviously, $T^H\subseteq E(I)$,
since $T^H=E(H)$. Moreover, since every vertex in $X(I)$ is adjacent to
both $u$ and $v$ in $f(I)$, $T^X\subseteq E(I)$. Finally, for every clause
$c$ in $C(I)$, edge set $T^c$ is defined as a subset of $E(I)$, so
$T^Q\subseteq E(I)$. Therefore, $T$ is a subgraph of $f(I)$; part of such a
graph $T$ is shown in figure~\ref{fNP4diam5} (a).
Edge $uv$ is in $T$ and let $K$ be the subgraph of $T$ induced by the
pair $u$ and $v$.

Let $c$ be an arbitrary clause in $C(I)$, which contains variables $x_1$,
$x_2$ and $x_3$, say. Then, $x(c)=x_i$ for some $i$ from 1 to 3.
Vertices $q_1^c,\ldots,\,q_8^c$ of $V^c$ are adjacent in $T$ to $x_i$ and
$x_i^c$ only, because edges of $T^c$ incident to
$q_1^c,\ldots,\,q_8^c$ are exactly the edges of $f(I)$ from $\{x_i,\,x_i^c\}$
to $\{q_1^c,\ldots,\,q_8^c\}$. Moreover, for $1\leq j \leq 8$, vertex $q_j^c$
has degree 1 in $T$, because exactly one of $M_{(2i-1),\,j}$ and $M_{2i,\,j}$
is 1; observe that the $(2i-1)$ row of $M$ is the complement of the $2i$ row
of $M$. On one hand, if variable $x_i$ appears positive in $c$, then $x_i^c$
is adjacent to $u$ in $T$ and $x_i$ is adjacent to $u$ in $T$, since $x_i$ is
a variable of $c$ that makes $c$ true with respect to truth assignment $A$.
On the other hand, if the negation of $x_i$ appears in $c$ then $x_i^c$
is adjacent to $v$ in $T$ and $x_i$ is adjacent to $v$ in $T$, since $x_i$
is a variable of $c$ that makes $c$ true with respect to truth assignment $A$.
So, both $x_i$ and $x_i^c$ are adjacent to the same vertex of $K$ in $T$.
Therefore, the distance in $T$ between any pair of vertices in
$\{q_1^c,\ldots,\,q_8^c\}$ is at most $4$ and all the vertices in
$\{q_1^c,\ldots,\,q_8^c\}$ are at distance 2 in $T$ from $K$.
The remaining vertices of $V^c$, i.e. vertices
$x_j^c$ where $j\not=i$ and $1\leq j \leq 3$, are adjacent in $T$ to
exactly one vertex of $K$ and have degree 1 in $T$.

Each vertex
in $X(I)$ is adjacent to exactly one vertex of $K$ in $T$. Vertices
in $H$ form a path in $T$ and every vertex in $H$ is at distance
at most 2 in $T$ from $K$. So, each vertex in $V(I)=Q(I)\cup X(I)\cup H$
is either adjacent to some vertex in $K$ or at distance 2 from $K$, thus
$T$ is connected and $K$ is a 5-center of $T$, i.e. $T$ has diameter at
most 5 (lemma~\ref{ldiametros}).

Next, we prove that $T$ does not contain any cycles. A vertex in $T$ but not
in $K$ has degree more than 1 in $T$ only if it is in $X(I)$ or it is
vertex $x(c)$ for some clause $c$ in $C(I)$ or it is vertex
$h_u'$ or $h_v'$; but all these vertices are adjacent in $T$ to exactly
one vertex of $K$
and there is no edge of $T$ between them. So, the vertices in $T$ of
degree more than 1 form a subtree of $T$, which implies that $T$ does
not have any cycles. Hence, $T$ is a tree.

Finally, in order to prove that $T$ is a $4$-spanner of $f(I)$, it suffices
to examine vertices at distance 2 from $K$ in $T$, since the remaining
vertices of $f(I)$ are adjacent to $K$ in $T$ and, therefore, they are within
distance 4 in $T$ from every vertex of $f(I)$, because $K$ is a 5-center of
$T$. The vertices of $f(I)$ that are at distance 2 from $K$ in $T$ are
vertices $h_u$, $h_v$ and vertices $q_1^c,\ldots,\,q_8^c$, for each clause
$c$ of $C(I)$. Obviously, each of $h_u$ or $h_v$ is adjacent in $T$ to its only
neighbor in $f(I)$, namely $h_u'$ or $h_v'$, respectively. For every pair of
clauses $c$ and $p\,$ of $C(I)$, there is no edge of $f(I)$ between a vertex in
$\{q_1^c,\ldots,\,q_8^c\}$ and a vertex in $\{q_1^p,\ldots,\,q_8^p\}$. Also,
for every clause $c$ in $C(I)$, the vertices in $\{q_1^c,\ldots,\,q_8^c\}$ are
at distance at most 4 in $T$ from each other, as we pointed out earlier in
the proof. Hence, $T$ is a tree 4-spanner of $f(I)$ of diameter at most 5.

Let $T'$ be a tree 4-spanner of $f(I)$ of diameter at most 5. Only vertices
$h_u'$ and $u$ are at distance at most 2 from $h_u$ and only
vertices $h_v'$ and $v$ are at distance at most 2 from $h_v$, so
pair $u$ and $v$ is the only 5-center $K$ of $T'$. Since $f(I)$
admits a tree 4-spanner with 5-center $K$, $f(I)$ admits a tree
4-spanner $T$ with 5-center $K$, such that $d_{f(I)}(x,K)=d_T(x,K)$, for
every vertex $x$ in $f(I)$, because of theorem~\ref{tisesapostaseis}. Every
vertex in $X(I)$ is adjacent to both $u$ and $v$ in $f(I)$, so,
for every $x$ in $X(I)$, at least one of edges $xu$ or $xv$ is in $T$, since
$T$ is a shortest paths to $K$ spanning tree of $f(I)$. Edge $uv$ is in
$T$ ($K$ is a 5-center of $T$), so not both edges $xu$ and $xv$ are in
$T$, because $T$ does not contain any triangles. Hence, for every vertex
$x$ in $X(I)$, exactly one of edges $xu$ or $xv$ is in $T$. Therefore, the
following definition of $A$ is a truth assignment to variables in $X(I)$:
for every $x$ in $X(I)$, $A(x)$ is true, when edge $ux$ is in $T$, and $A(x)$
is false, when edge $vx$ is in $T$. We prove that $A(x)$ satisfies $I$. 

Let $c$ be an arbitrary clause
in $C(I)$, which contains variables $x_1$, $x_2$
and $x_3$, say. Every vertex in $\{q_1^c,\ldots,\,q_8^c\}$
is at distance exactly 2 from $K$ in $T$, because $K$ is a 5-center of $T$
and none of $q_1^c,\ldots,\,q_8^c$ is adjacent to $K$ in $f(I)$. So, each
of $q_1^c,\ldots,\,q_8^c$ has to be adjacent to at least one of the vertices
in $g^c$, because vertices in $g^c$ are the only neighbors in $f(I)$ of
vertices in $\{q_1^c,\ldots,\,q_8^c\}$ that are adjacent to a vertex in $K$.
If we pick one vertex from each of the following three pairs $x_1$ and
$x_1^c$, $x_2$ and $x_2^c$, $x_3$ and $x_3^c$, then the neighborhood in $f(I)$
of this triplet of vertices does not include all the vertices in
$\{q_1^c,\ldots,\,q_8^c\}$; observe that, because of the structure of
matrix $M$, unless we pick two complementary rows of $M$ we cannot have
a (proper) submatrix of $M$ consisting of rows of $M$, such that the submatrix
has at least one 1 in each column (see figure~\ref{fNP4diam5} (b) for an
example). Therefore, for at least one $i$ from 1 to 3,
both of $x_i$ and $x_i^c$ are adjacent to some vertices in
$\{q_1^c,\ldots,\,q_8^c\}$. Also, towards a contradiction, assume that
$x_i^c$ is adjacent to $u$ and $x_i$ to $v$ in $T$ or $x_i^c$ to $v$ and
$x_i$ to $u$, i.e. assume that vertices $x_i$ and $x_i^c$ are adjacent to
different vertices of $K$ in $T$. Then, $K$ is the 4-midst of a path $P$
in $T$ with endpoints in $\{q_1^c,\ldots,\,q_8^c\}$. But, there is another
path in $f(I)$ that avoids the neighborhood of $K$ in $f(I)$
between the two endpoints of $P$ (recall that the vertices in
$\{q_1^c,\ldots,\,q_8^c\}$ induce a path in $f(I)$),
which is a contradiction, because of lemma~\ref{lofthalmos}. Thus,
both of $x_i$ and $x_i^c$ are adjacent in $T$ to the same vertex of $K$.

On one hand, if variable $x_i$ appears positive in $c$, then
$x_i^c$ is adjacent to $u$, because of the construction of graph $f(I)$.
So, since $x_i$ and $x_i^c$ are adjacent to the same vertex of  
$K$ in $T$, vertex $x_i$ is also adjacent to $u$ in $T$. So, in this case,
$A(x_i)$ is true by the definition of $A$. Therefore, clause $c$ is
satisfied by truth assignment $A$. On the other hand, if the negation
of $x_i$ appears in $c$, then, similarly, $A(x_i)$ is false; so, clause
$c$ is satisfied by truth assignment $A$. In both cases $A$ satisfies
the arbitrary clause $c$ in $C(I)$. Also, if $c$ is not in $C(I)$ then $c$
contains 3 literals but 2 variables, i.e. $c$ contains both a variable and
its negation, so $A$ satisfies $c$. Hence, truth assignment $A$
satisfies all clauses of $I$. \myqed

Therefore, the problem of determining if a graph admits a tree 4-spanner
of diameter at most 5 is an NP-complete problem.

\subsection{The remaining values}
In this section, $t\geq 5$. Given graph $f(I)$, where $I$ is an instance of
3-SAT, we describe a second  graph $h(f(I),t)$ and, furthermore, we show
that the second
graph admits a tree $t$-spanner of diameter at most $t+1$ if and only if
the first graph admits a tree 4-spanner of diameter at most
5. Therefore, for each $t\geq 5$ and for every instance $I$ of 3-SAT,
graph $h(f(I),t)$ admits a tree $t$-spanner of diameter at most $t+1$
if and only if $I$ is satisfiable, because of proposition~\ref{pNP4diam5}.
Hence, for each $t\geq 5$, the problem of determining if a graph admits
a tree $t$-spanner of diameter at most $t+1$ is an NP-complete problem.
The main gadget for this reduction is to add to $f(I)$ a path of length $t-3$
with endpoints the two vertices of the 5-center of $f(I)$ so that paths
of length 5 in a tree 4-spanner of $f(I)$ become paths of length $t+1$ in
a tree $t$-spanner of graph $h(f(I),t)$.

\begin{figure}[htbp]
\begin{center}
\includegraphics{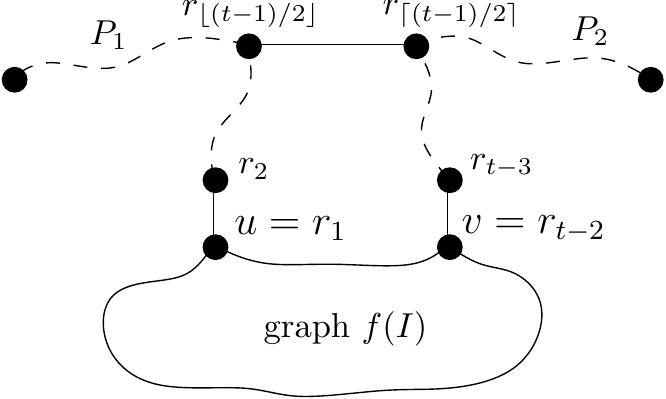}
\caption{Graph $h(f(I),t)$, where $t$ is even. Here, its subgraph $f(I)$ is not shown in detail.}
\label{ftspandiamt+1}
\end{center}
\end{figure}

More formally, consider such a graph $f(I)$. For each $t\geq 5$, consider the
following graph $R(t)$, for which $f(I)\cap R(t)=\{u,\,v\}$.
First, let $P$ be a path $u$=$r_1,\,\ldots,\,r_{t-2}$=$v$ of length $t-3$.
Second, let $P_1$  be a path of length $\lfloor \frac{t+1}{2}\rfloor$
having vertex $r_{\lfloor (t-1)/2\rfloor}$ of $P$ as an endpoint and no other
vertex in common with $P$. Third, let $P_2$ be a path of length
$\lfloor \frac{t+1}{2}\rfloor$ having vertex $r_{\lceil (t-1)/2\rceil}$ of $P$ 
as an endpoint and no other vertex in common with $P$ or $P_1$. Note that,
when $t$ is odd, paths $P_1$ and $P_2$ share an endpoint, namely vertex
$r_{\lfloor (t-1)/2\rfloor}$=$r_{\lceil (t-1)/2\rceil}$.
Now, $R(t)$ is the union of $P$, $P_1$ and $P_2$.
Graph $h(f(I),t)$ is defined as $R(t)\cup f(I)$ (figure~\ref{ftspandiamt+1}).
For each $t\geq5$, note that, given $I$, graph $h(f(I),t)$ can be constructed in polynomial time.

\begin{prop}
For each $t\geq 5$,
for every instance $I$ of 3-SAT, graph $f(I)$ admits a tree 4-spanner of
diameter at most 5 if and only if graph $h(f(I),t)$ admits a tree
$t$-spanner of diameter at most $t+1$.
\end{prop}
{\em Proof.}
Assume that graph $f(I)$ admits a tree 4-spanner $T$ of diameter 5.
Then, the pair of vertices $u$ and $v$ is the only 5-center of $T$. We prove
that the subgraph $T'$ of $h(f(I),t)$ being the union of $R(t)$ and
$(T\setminus\{uv\})$ is a tree $t$-spanner of $h(f(I),t)$ of diameter at most $t+1$.
Obviously, $T'$ is a spanning tree of $h(f(I),t)$, since $R(t)$ is a
spanning tree of itself, $T$ is a spanning tree of $f(I)$ and edge $uv$ is
not in $T'$. Let $K'$ be the subgraph of $T'$ induced by the pair of vertices
$r_{\lfloor (t-1)/2\rfloor}$ and $r_{\lceil (t-1)/2\rceil}$. Note that, when
$t$ is odd, $K'$ consists of one vertex and, when $t$ is even, $K'$ consists
of a pair of adjacent in $T'$ vertices. Every vertex
of $R(t)$ is within distance $\lfloor\frac{t+1}{2}\rfloor$ from $K'$ in
$T'$, since paths $P_1$ and $P_2$ have length $\lfloor\frac{t+1}{2}\rfloor$
and $K'$ lies in the middle of path $P$ which has length $t-3$.
Also, every vertex of $f(I)$ is within distance 2 in $T'$ from at least
one of $u$ and $v$, since $u$ and $v$ induce a 5-center of $T$
(note that edge $uv$ is not in $T'$), where each of
$u$ and $v$ is at distance $\lfloor \frac{t-3}{2}\rfloor$ from $K'$ in $T'$;
so, every vertex of $f(I)$ is within distance
$\lfloor\frac{t+1}{2}\rfloor$ from $K'$ in $T'$. Therefore, $K'$ is
a $(t+1)$-center of $T'$, i.e. $T'$ has diameter at most $t+1$
(lemma~\ref{ldiametros}).

Next, we prove that $T'$ is a tree $t$-spanner of $h(f(I),t)$. Since all
edges of $R(t)$ are in $T'$, it suffices to examine the adjacencies between
vertices in $f(I)$. If two vertices of $f(I)$ are adjacent in $h(f(I),t)$,
then they are within distance 4 from each other in $T$, since $T$ is a
tree 4-spanner of $f(I)$.
But, $d_{T'}(x,y)\leq d_T(x,y)-1+t-3=d_T(x,y)+t-4$, for every pair of vertices $x$ and
$y$ in $f(I)$, since $T'$ contains path $P$ of length $t-3$ instead of
edge $uv$ of $T$ (of course, when the $x,\,y$-path of $T$ does not contain
edge $uv$, $d_{T'}(x,y)=d_T(x,y)$). So, every pair of adjacent in
$f(I)$ vertices are within distance $t$ from each other in $T'$
(note that if $x$ and $y$ are adjacent in $f(I)$, then $d_T(x,y)\leq 4$).
Hence, $T'$ is a tree $t$-spanner of $h(f(I),t)$ of diameter at most $t+1$.

Assume that graph $h(f(I),t)$ admits a tree $t$-spanner of diameter at most
$t+1$. Vertices $r_{\lfloor (t-1)/2\rfloor}$ and $r_{\lceil (t-1)/2\rceil}$
of $R(t)$ induce the only $(t+1)$-center $K'$ of this tree $t$-spanner of
$h(f(I),t)$, because of paths $P_1$ and $P_2$ of $R(t)$. Thus, because of
theorem~\ref{tisesapostaseis}, graph $h(f(I),t)$ admits a tree $t$-spanner
$T'$ which is a shortest paths to $K'$ spanning tree of $h(f(I),t)$.
For each of $u$ and $v$ there is a unique shortest path in $h(f(I),t)$ to
$K'$; so path $P$ of $R(t)$ is a subpath of $T'$ and, therefore, edge $uv$
is not in $T'$.
We prove that $T=T'[f(I)]\cup\{uv\}$ is a tree 4-spanner of $f(I)$ of diameter
at most 5, where $T'[f(I)]$ is the subgraph of $T'$ induced by the vertices
of $f(I)$. Obviously, $T$ is a spanning tree of $f(I)$, since $T$ does
contain edge $uv$, instead of path $P$ of $T'$.

Let $x$ be an arbitrary vertex of $f(I)$. The shortest path in $T'$ from
$x$ to $K'$ consists of either the $x,\,u$-path of $T$ and the $u,\,K'$-path
of $T'$ or the $x,\,v$-path of $T$ and the $v,\,K'$-path of $T'$.
But, each of the $u,\,K'$-path and the $v,\,K'$-path has length
$\lfloor \frac{t-3}{2}\rfloor$ and, furthermore, $x$ is within distance
$\lfloor\frac{t+1}{2}\rfloor$ from $K'$ in $T'$. So, vertex $x$ of $f(I)$
is within distance 2 in $T$ from at least one of $u$ or $v$. Therefore,
since edge $uv$ is in $T$, the pair $u$ and $v$ induces a 5-center
$K$ of $T$; i.e. T has diameter at most 5 (lemma~\ref{ldiametros}).

In order to prove that $T$ is a tree 4-spanner of $f(I)$, it suffices
to examine vertices at distance 5 from each other in $T$, since $T$ has
diameter at most 5. For any two vertices $x$ and $y$ of $f(I)$ at distance
5 from each other in $T$, the path of $T$ between $x$ and $y$ contains
edge $uv$, since $K$ is a 5-center of $T$. Therefore, $d_{T'}(x,y)=$
$5-1+t-3=t+1$, since $T'$ does not contain edge $uv$ but contains path $P$
of $R(t)$ which has length $t-3$. Thus, $x$ and $y$ are not adjacent to
each other in $f(I)$, since $T'$ is a tree $t$-spanner of $h(f(I),t)$.
Hence, $T$ is a tree 4-spanner of $f(I)$. \myqed

\section{The situation for each value of the stretch factor}
When $t\leq 1$, the tree $t$-spanner problem is trivial. Also,
one can prove that there is an efficient algorithm to determine if
a graph admits a tree $2$-spanner of diameter at most $3$. The following
theorem summarizes the results in this chapter.

\begin{thm}
For each $t$, the problem of deciding if a graph admits a tree $t$-spanner
of diameter at most $t+1$ can be solved efficiently, when $t\leq 3$, and it is
an NP-complete problem, when $t\geq 4$.
\label{tgiaolatat}
\end{thm}

\section{Acknowledgments}
Results in this paper appear in my PhD thesis \cite{PhDthesis} at the department of computer science of the university of Toronto;
I would like to thank my thesis supervisor, professor Derek Corneil.

\bibliographystyle{plain}
\bibliography{tspanners}
\end{document}